\documentclass[aps,pre, twocolumn, amsmath,amssymb,floatfix, showpacs]{revtex4-1}
\usepackage{graphicx}% Include figure files
\usepackage{dcolumn} % Align table columns on decimal point
\usepackage{bm}      % Bold math
\usepackage[usenames,dvipsnames]{color}
\usepackage{ulem} % for strikeout font (\sout)
 \usepackage{csquotes} \usepackage{tikz} \usetikzlibrary{shadings,patterns} \usepackage[capitalize]{cleveref}

\begin{document}

\title{Geophysical turbulence and the duality of the energy flow across scales}

\author{ A. Pouquet$^{\ast 1,2}$ and R. Marino$^1$}
\affiliation{
 $^1$Computational and Information Systems Laboratory, NCAR,  
 Boulder CO 80307, USA. \\
$^2$Department of Applied Mathematics, University of Colorado at Boulder, Boulder, CO 80309, USA. 
                 }

 \begin{abstract}
The ocean and the atmosphere, and hence the climate, are  governed at large scale by interactions between  pressure gradient, Coriolis  and  buoyancy forces. This leads to a quasi-geostrophic balance in which, in a two-dimensional-like fashion, the energy injected by solar radiation, winds or tides goes to large scales in what is known as an inverse cascade. Yet, except for Ekman friction, energy dissipation and turbulent mixing occur at small scale implying the formation of such  scales 
%in a direct energy cascade 
associated with breaking of geostrophic dynamics through wave-eddy interactions \cite{ledwell_00, vanneste_13} or  frontogenesis \cite{hoskins_72, mcwilliams_10}, in opposition to the inverse cascade. Can it be both at the same time? % How do these phenomena co-exist? 
%
%%%There are idealized physical systems,  modeling complex fluids under rather restrictive conditions, % as occur in geophysics and astrophysics, 
%%% that exhibit such a dual behavior of energy flowing to the large and to the small scales, with constant fluxes as required by theoretical arguments. 
%
%We appeal to phenomenological models and simplified representations of turbulence to support this concept, and w
We exemplify here this dual behavior of energy with the help of three-dimensional direct numerical simulations of rotating stratified Boussinesq turbulence. %, indicating that 
We show that  efficient small-scale mixing and large-scale coherence develop %inverse flux do  occur 
simultaneously in such geophysical and astrophysical flows, both with constant flux as required by theoretical arguments, thereby clearly resolving the aforementioned contradiction. 
 \end{abstract}

	% Direct numerical simulations
				% 47.35.Bb,  % Hydrodynamic waves (fluids)
		
\maketitle
Geostrophic balance, in which nonlinearities are neglected, leads to simplified quasi-bi-dimensional behavior with energy flowing to large scales, and reduced 
small-scale dissipation, contrary to observations \cite{ivey_08}: vertical mixing can decrease water density, 
contributing to the (upward) closing of the ocean global circulation \cite{ledwell_00}. It is identified with breaking of internal gravity waves 
\cite{nikurashin_13}, and it can potentially control the amplitude of the mesoscales.

Such flows are neither three-dimensional (3D) nor two-dimensional (2D), since at small scales, 3D eddies may prevail. Considering the system dimensionality 
$D_S$ proved essential when examining critical phenomena which simplify in higher dimensions, due to more  mode interactions  as $D_S$ grows.  
Fluid turbulence is  
vastly different in two or three dimensions, because of the strong  constraint imposed by the new 2D invariants (such as the integrated powers of vorticity). 
This leads to energy flowing towards the largest scales, ending up in a condensate \cite{rhk_montgo}; it  
can take the form of 
features such as jets, observed in the atmosphere of planets, or in the oceans as  
striations \cite{galperin_04}. Thus, geophysical turbulence is anisotropic, quasi-2D at large scale and quasi-3D at small scale \cite{sagaut_cambon_08}.

However, traditional three-dimensional homogeneous isotropic turbulence (HIT) is known to break structures (meso-scale eddies, clouds)
into progressively smaller entities which will be dissipated at small scale, enhancing mixing of tracers such as pollutants \cite{shraiman_00} or biota \cite
{klein_09}. Whereas the fate of energy in 3D is modeled through an enhanced viscosity $\nu_{turb}>0$, the 2D evolution leading to large-scale structures can be  
related to a destabilizing transport coefficient, e.g.  $\nu_{turb}\le 0$.
Since the direction of the cascade is known to affect the amount of energy available to irreversible processes of dissipation and mixing, it is thus an essential
parameter in the overall energy budget of the atmosphere and ocean \cite{ferrari_09}.
 
A transition from 2D to 3D in turbulence has been investigated in various contexts. For example, is there a critical dimension for which $\nu_{turb}$ changes
sign, indicative of a change of behavior in the overall flow dynamics? Using two-layer quasi-geostrophic (QG) models with bottom friction, it was shown recently 
that when adding, in a somewhat ad-hoc fashion, a horizontal eddy-viscosity mimicking coupling to smaller scales and thereby presumably changing locally the sign of $\nu_{turb}$, both a direct and inverse energy cascades {were} obtained \cite{arbic_13}.  

More formally, starting from two-point turbulence closure, space dimensionality appears % as a parameter when using 
through incompressibility. The critical dimension that separates 2D from 3D behavior can  
be computed and is found to be $\approx 2.05$  \cite{frisch_76} (see also \cite{fournier_78}). 
 A simple model which is a local version (in modal space) of the closure equations, derived in 
\cite{bell_77}, describes the energy flux to the small scales and the large scales by introducing an (unsigned) parameter which represents the ratio of inverse
to direct flux, 
$$R_\Pi=|\epsilon_{I}/\epsilon_{D}| \ ; $$ 
it is found to be a smooth monotonic function of 
 $D_S$, in a fashion similar to critical phenomena, thus providing a path between 2D and 3D behavior.
In order to model the  anisotropy of geophysical flows, one can alternatively introduce an anisotropic scale contraction/dilation. This allows to break the 
geostrophy constraint by considering explicitly the production of horizontal vorticity by horizontal or vertical eddies; it leads to a fractal dimension of
turbulence, close to 2.55 for stratified flows \cite{lovejoy_12a}.  

Furthermore, an inverse energy cascade  can also occur in 3D-HIT. On the one hand, when restricting nonlinear interactions in 3D to those between helical waves
of the same polarization, energy is found to flow to large scale, with helicity (velocity-vorticity correlations) populating the small scales 
\cite{biferale_13a}. In reality, cross-polarization interactions dominate, but the tendency for strong inverse transfer is clearly displayed in this restricted model. 

On the other hand, taking a purely 2D input of energy and a fluid  with a variable aspect ratio $A_r$, energy again has an increased tendency to flow to large 
scales as $A_r$ becomes small, with a transition at $A_r\approx 1/2$ ($A_r$ is defined as the ratio of the vertical resolution to the forcing scale)  \cite
{celani}. A clear dual energy cascade obtains, with $R_\Pi$  a decreasing function of $A_r$. Also, inverse transfer in thick layers (with now $A_r \approx 0.78$) 
is observed experimentally, the suppression of vertical motions being attributed to interactions with vertical shear for eddies whose time-scale is larger than 
the characteristic shear time \cite{xia_11}.

These are idealized physical systems,  modeling complex fluids under rather restrictive conditions.
However, the link between large scales and small scales (or nonlocal interactions between Fourier modes) is embodied in coherent structures such as  chlorophyl  filaments \cite{davis_04}, water vapor, ozone, 
temperature or salinity tracer fronts, and in magnetohydrodynamics, current sheets, plasmoids and Alfv\'en vortices \cite{sundkvist_05}. These structures have one dimension comparable to the integral scale of the flow or larger, and one close to the dissipative scale. 
One element altering the way such structures arise and evolve is the ideal invariants, and in particular whether or not they involve gradients. 
Finally, if one expects the symmetries of the primitive equations to recover at small scale, using a statistical argument based on the large number of modes, 
this recovery may be impeded by the presence of large-scale  shear \cite{pumir_95}.
For example, direct coupling between large scales (at which the inertio-gravity waves reside) and small scales (at which turbulence resides) was demonstrated in 
\cite{fritts_09a}, providing a progressive destruction of shear layers together with propagation, over the layer depth, of efficient mixing induced by the 
turbulence.

Stratified turbulence is not 2D in the traditional sense: it has strong vertical shearing 
\cite{chomaz, lindborg2006, brethouwer_07, waite_08, sagaut_cambon_08, billant_10}, 
% \cite{chomaz, brethouwer_07, sagaut_cambon_08},  NOTE: We could regroup the others in a traditional system ... Will do if space needed in revised version
allowing for the efficient creation of small scales, as well as of large scales in the presence of rotation \cite{EPL}. 
What is perhaps not well recognized is that the {3D} Boussinesq equations, including rotation and stratification as in the 
atmosphere and oceans, can produce both large scale and small scale energy excitation, {\sl both} with constant flux. Numerous numerical studies suffer from a 
lack of resolving both the large and the small eddies: because of the inherent cost of such computations, a divide-and-conquer approach has been successfully 
followed, analyzing either the direct or the inverse cascade, but not convincingly both. {Fluxes of energy to large scales and to small scales} become comparable 
for strong rotation \cite{3072}, as well as in the presence of stratification \cite{aluie_11}.
However, in all these studies, the smallness of the forcing wavenumber ($\approx 4$ or $5$) does not allow for a clear conclusion concerning the existence of the 
inverse cascade itself.

\begin{table} \begin{ruledtabular}    \begin{tabular}{c|c|c|c|c|c|c|c}
Run  &  $Re$  &  $Fr$   &  $Ro$  &  N/f  &  ${\cal R}_B$  &  $R_\Pi$  &  $\alpha$ \\
\hline
10a  &  5000  &  0.020  &  0.08  &   4   &      2.0     &    5.77   &   -3.99    \\
10b  &  5000  &  0.045  &  0.18  &   4   &     10.1     &    2.70   &   -2.93     \\
10c  &  5000  &  0.060  &  0.24  &   4   &     18.0     &    1.36   &   -2.34     \\
\hline
10d  &  4000  &  0.040  &  0.08  &   2   &      6.4     &    9.04   &   -3.99     \\
10e  &  5000  &  0.090  &  0.18  &   2   &     40.5     &    1.62   &   -2.12     \\
\hline
15a  &  8000  &  0.100  &  0.20  &   2   &     80.0     &    1.08   &   -1.87     \\
\end{tabular}  \end{ruledtabular} \label{tab}

\caption{
List of the runs done on cubic grids of $n_p^3$ points, with 10 \& 15 %in the ``Run'' column 
standing for $n_p=1024$ \& $1536$ respectively. All runs use a random force in the wavenumber band % centered on 
$k_F\in [10,11]$. $Re, \ Fr$ and $Ro$ are the Reynolds, Froude and Rossby numbers, with $N/f=Ro/Fr$ and ${\cal R}_B=ReFr^2$ the buoyancy Reynolds number. 
$R_\Pi=\epsilon_{I}/\epsilon_{D}$ is % the absolute value of 
the ratio of the direct to the inverse flux of energy in the vicinity of $k_F$ % the forcing band 
(1 $<$ k $<$ 9 for $\epsilon_{I}$, $11 < k < 20$ for $\epsilon_{D}$); it is computed on spectra averaged over 10 turn-over times $\tau_{NL}=L_F/U_0$, in the range $ 12 < t/{{\tau}_{NL}} < 22$.
Finally, $\alpha$ is the best fit for the small-scale kinetic energy spectral index; note the significant decrease of $\alpha$ with increasing {Re and} ${\cal R}_B$. All large-scale indices, computed for $k<k_F$, are close to $5/3$ (see Fig. \ref{2runs}). 
} \end{table}

\begin{figure}[h!tbp] \centering
\resizebox{8.5cm}{!}{\includegraphics{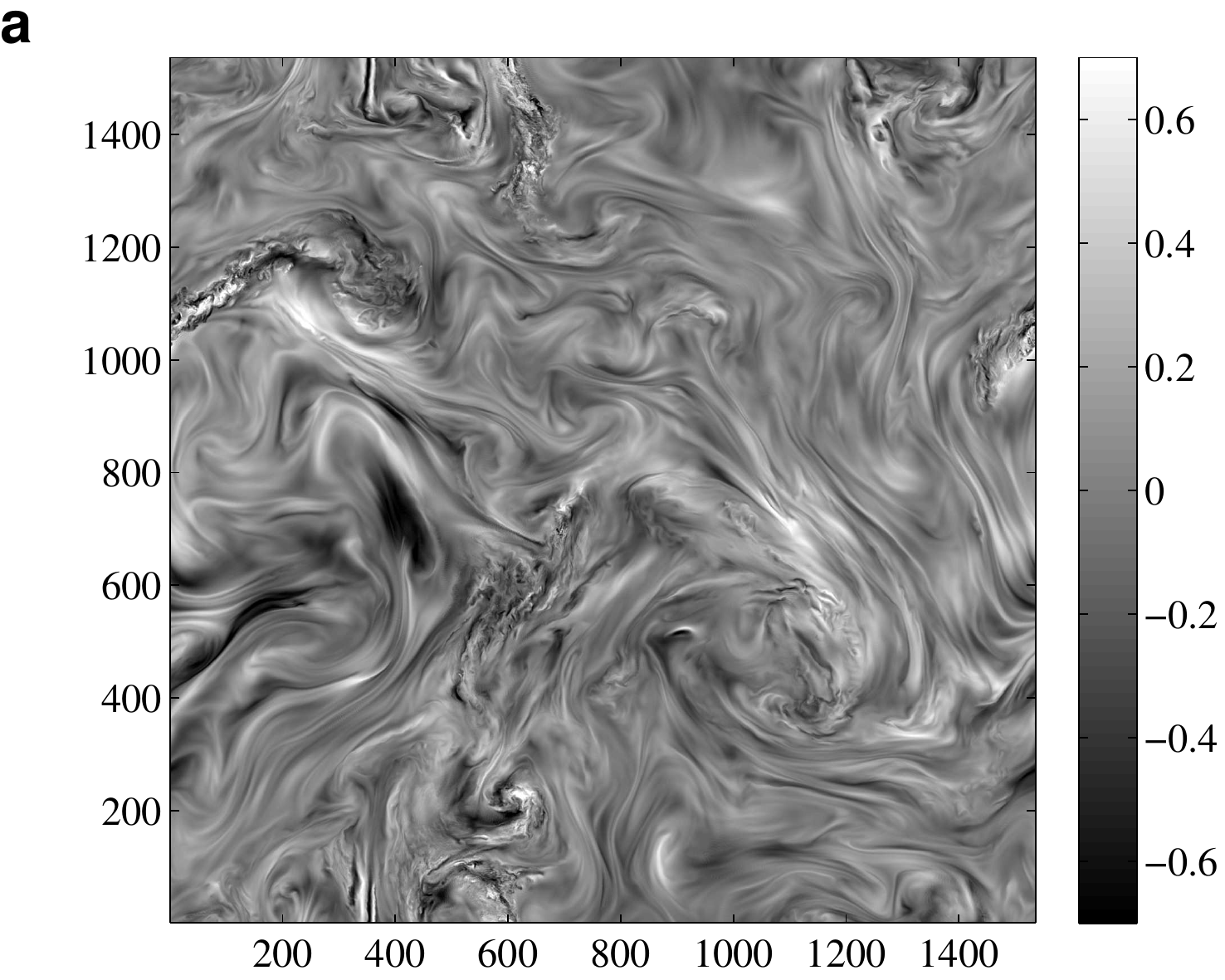}}
\vskip-0.14truein
\resizebox{8.5cm}{!}{\includegraphics{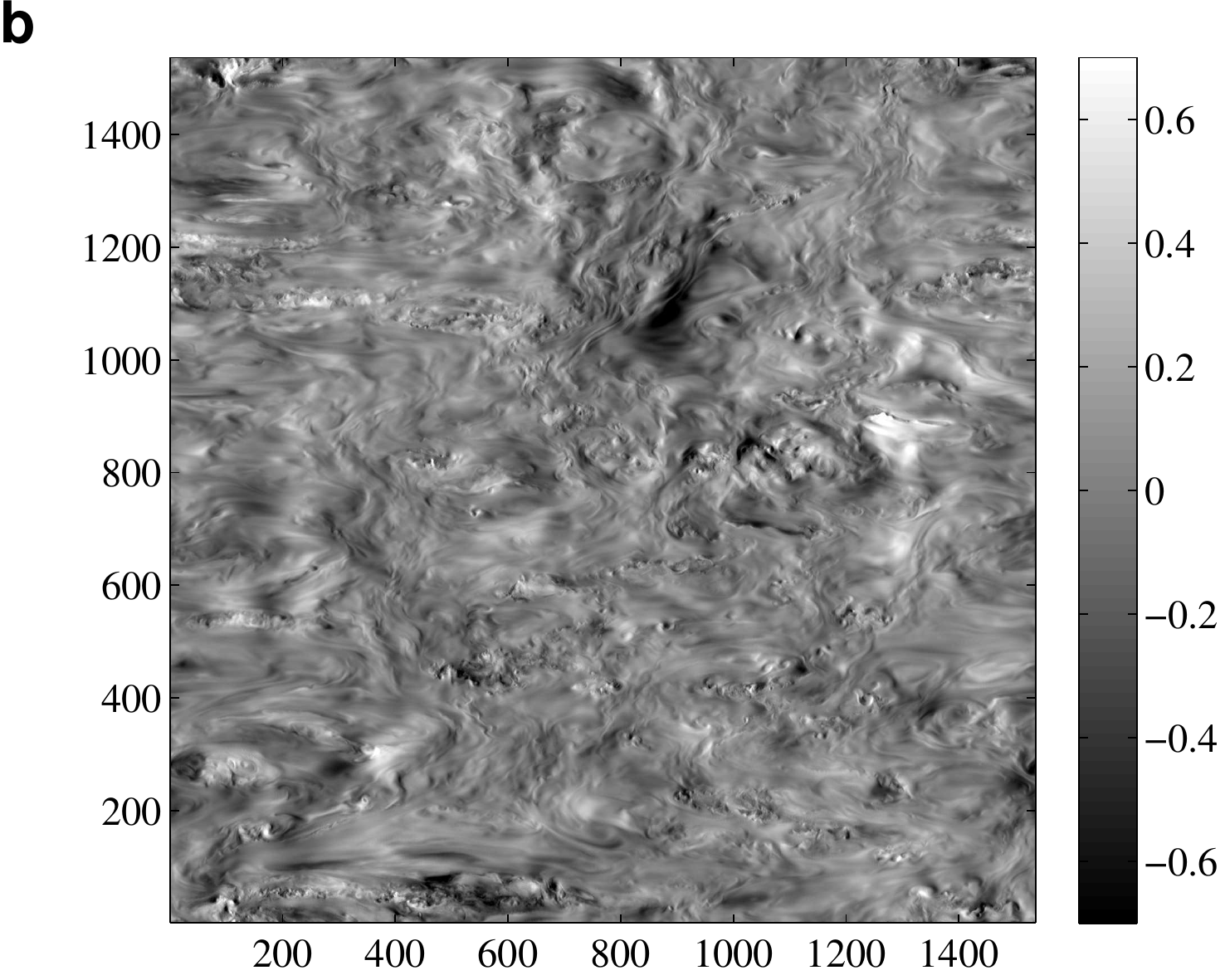}}
\vskip-0.14truein

\caption{\label{cuts}  
Horizontal (xy, {\bf (a)}) and  vertical (xz, {\bf (b)}) two-dimensional cuts of the vertical velocity for run 15a at the latest time, 
with ${\cal R}_B\approx 80$ and a small-scale spectrum slightly steeper than a Kolmogorov law. 
The axes are labeled in terms of grid spacing, and the forcing scale corresponds to roughly 145 in these units. Observe the large-scale structures, with a size 
of up to a third of the overall flow (or more in the filaments), arising from the inverse cascade, together with super-imposed intense small scale eddies 
(e.g., at $x\approx 200, \ y\approx 1100$ in the top figure). Notice also the different structures in the two cuts, indicative of the persistent anisotropy of the flow.
}\end{figure} 

We thus now bring numerical evidence of the simultaneous generation of  large-scale and  small-scale flows, both with constant flux, using % large 
direct numerical simulations (DNS) of the  Boussinesq equations (see Table I). % \ref{tab}  %% and the Methods section below for details).
% These runs clearly point out to the possibility of a co-existence  in the ocean of  idealized large-scale dynamics dominated by quasi-geostrophic motions, together with the production of  small scales,  essential to mixing \cite{heywood_02}. 

{\underline{{\it Methods:}}}  
Oceanic turbulence is  studied in the idealized context of  the incompressible stably stratified rotating Boussinesq primitive equations, with ${\bf u}$ the velocity and $\theta$  the 
density fluctuations in units of velocity. Solid-body  rotation of strength $\Omega_0$  (with $f=2\Omega_0$) is imposed in the vertical ($z$) direction with unit vector $\hat z$, as well as anti-aligned gravity $g$; isotropic three-dimensional forcing ${\bf F}$ is included; $\nabla \cdot {\bf u}=0$ ensures incompressibility:
\begin{equation} 
 \partial_t {\mathbf u}   - \nu \Delta {\mathbf u} + N \theta \hat z 
+  {\bf F}+  \nabla p  - f {\bf u} \times \hat z = - {\mathbf u} \cdot \nabla {\mathbf u}  ,      \label{eq:mom} 
\end{equation} 
\begin{equation} 
 \partial _t \theta\   - \kappa \Delta \theta\ -   N w =  - {\mathbf u} \cdot \nabla \theta\    \  ,\ 
 \label{eq:temp} \end{equation} 
\noindent  
$w$ being the vertical  velocity, $p$ the pressure, $\nu$ the viscosity, and $\kappa=\nu$ the thermal diffusivity. 
 The square Brunt-V\"ais\"al\"a frequency is given by $N^2=-(g/\theta\ ) (d\bar \theta\ /dz)$, where $d\bar \theta/dz$ is the imposed background stratification,  assumed to be linear and constant.  In the ideal case ($\nu=0,\ {\bf F}=0$), the total (kinetic plus potential) 
energy $E_T= \frac{1}{2} \left<|{\bf u}|^2 + \theta\ ^2\right>=E_V+E_P$ is conserved 
and the point-wise potential vorticity 
$P_V= -N\omega_z +  f \partial_z \theta + \omega \cdot \nabla \theta$ 
is a material invariant.  No modeling of small-scale dynamics is included.  

 The numerical code, GHOST (Geophysical High Order Suite for Turbulence),  uses a pseudo-spectral method and is tri-periodic, with $n_p^3$ grid points; it is parallelized with a hybrid MPI/Open-MP method and scales linearly up to 98,000 processors for grid of up to $6144^3$ points \cite{hybrid2011}. 
Forcing is introduced in the momentum equation as a random field centered in the wavenumber band $k_F\in [10,11]$. The largest resolved scale is adimensionalized to $L_0=2\pi$, corresponding to a minimum wavenumber $k_{min}$=1; the smallest resolved scale is  $2\pi/k_{max}=6\pi/n_p$.
Initial conditions are zero for the density and random for ${\bf u}$. 
 
 Three dimensionless parameters characterize the flow:  the Reynolds number $Re=U_0L_F/\nu$, the Rossby number $Ro=U_0/[L_Ff]$ and  the Froude number, $Fr=U_0/[L_FN]$; 
 $U_{0}$ is the  $rms$ velocity, $L_F=2\pi/k_F$ is the forcing scale; 
 finally, $\epsilon_V\equiv dE_V/dt =- \left< {\bf u} \cdot {\bf F} \right>$ is the kinetic energy injection rate. 
Note that in order to resolve the Ozmidov scale, at which the eddy turn-over time and $1/N$ become equal and isotropisation recovers, one can show that ${\cal R}_B\ge 1$ where ${\cal R}_B=ReFr^2$ is the buoyancy Reynolds number.
Runs are performed with $2 < {\cal R}_B \le 120$ (see Table \ref{tab}). Whether the Ozmidov scale is properly resolved or not 
may well alter the efficiency of mixing, and the properties of stratified turbulence, as advocated in \cite{brethouwer_07} 
and as also observed here. 

The right-hand sides of equations (\ref{eq:mom}, \ref{eq:temp}) are used to derive the evolution of the total (kinetic + potential) energy density. Taking its Fourier transform (denoted by $\hat .$, with $\star$ denoting complex conjugate) gives access to the spectral transfer which, upon integration over wavenumber, yields the total isotropic energy flux $\Pi_T=\Pi_V+\Pi_P$: 
\begin{equation} 
\Pi_{V}(k)= \int_{k_{min}}^k T_{V}(q)dq \ , T_V(q)=  - \sum_{{\cal C}_q} {\hat{ {\bf u}}}^\star_{{\bf q}} \cdot \widehat{\left({\bf u} \cdot \nabla {\bf u}\right)}_{\bf q}
\nonumber    \end{equation} 
with ${\cal C}_q$ the shell $q\le |{\bf q}|< q+1$.  An expression for $\Pi_P$ can be written in a similar fashion.
 Note that in these Boussinesq runs, the  eventual change of sign of energy fluxes at a ``zero-crossing'' wavenumber is given by $k_F$ since the forcing is added at that scale.

{\underline{Results:}}
Fig.\ref{cuts} shows full 2D cuts of  vertical velocity in the vertical and horizontal for run 15a; % for the whole flow; 
the forcing is roughly 1/10th of the
box and one clearly observes both intense small-scale features where dissipation occurs, and organized patches significantly larger than the forcing scale,
indicative of the dual flux of energy.

\begin{figure*}
\vskip-0.57truein
\includegraphics[width=0.48\textwidth]{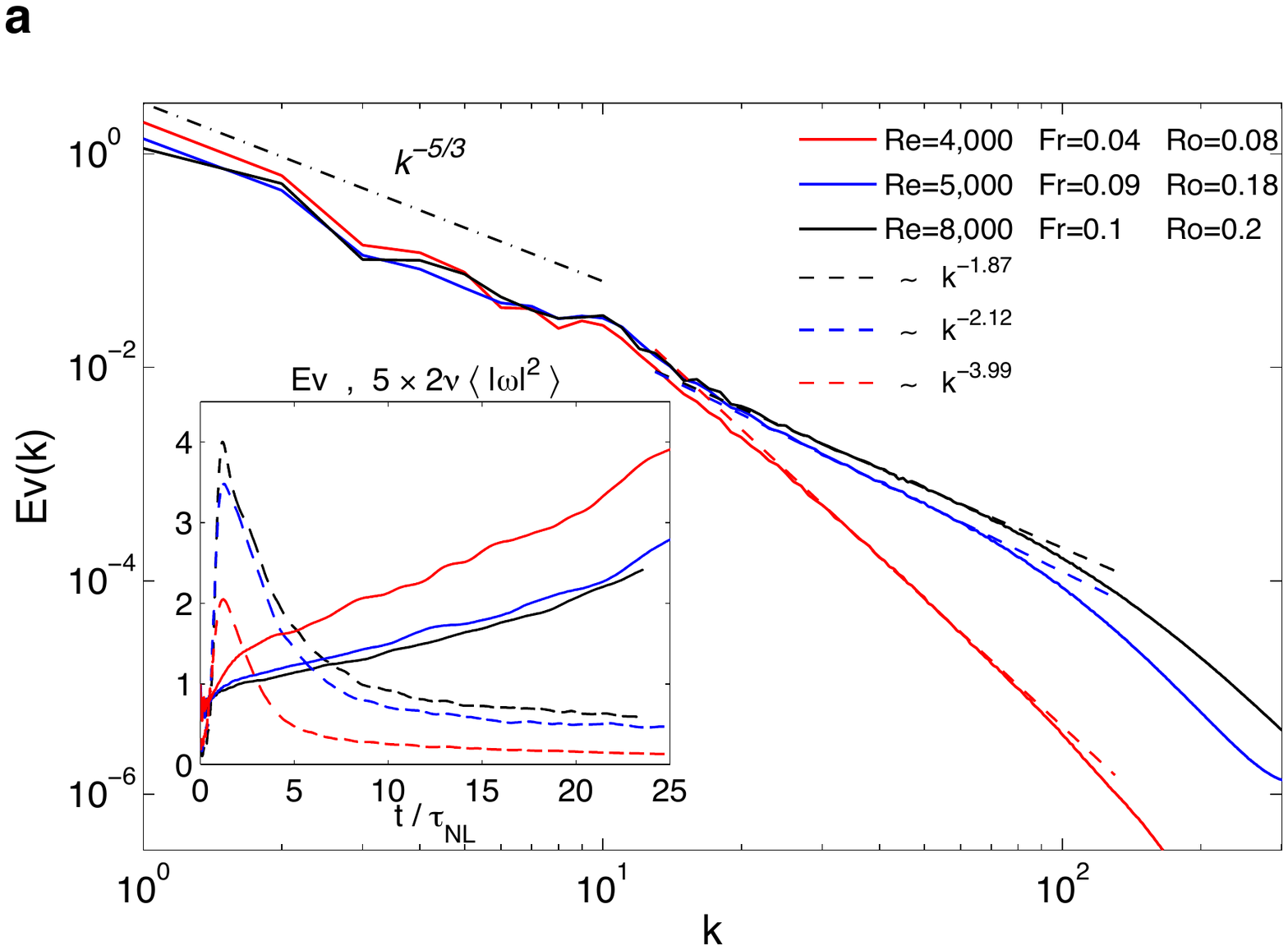} 
\includegraphics[width=0.48\textwidth]{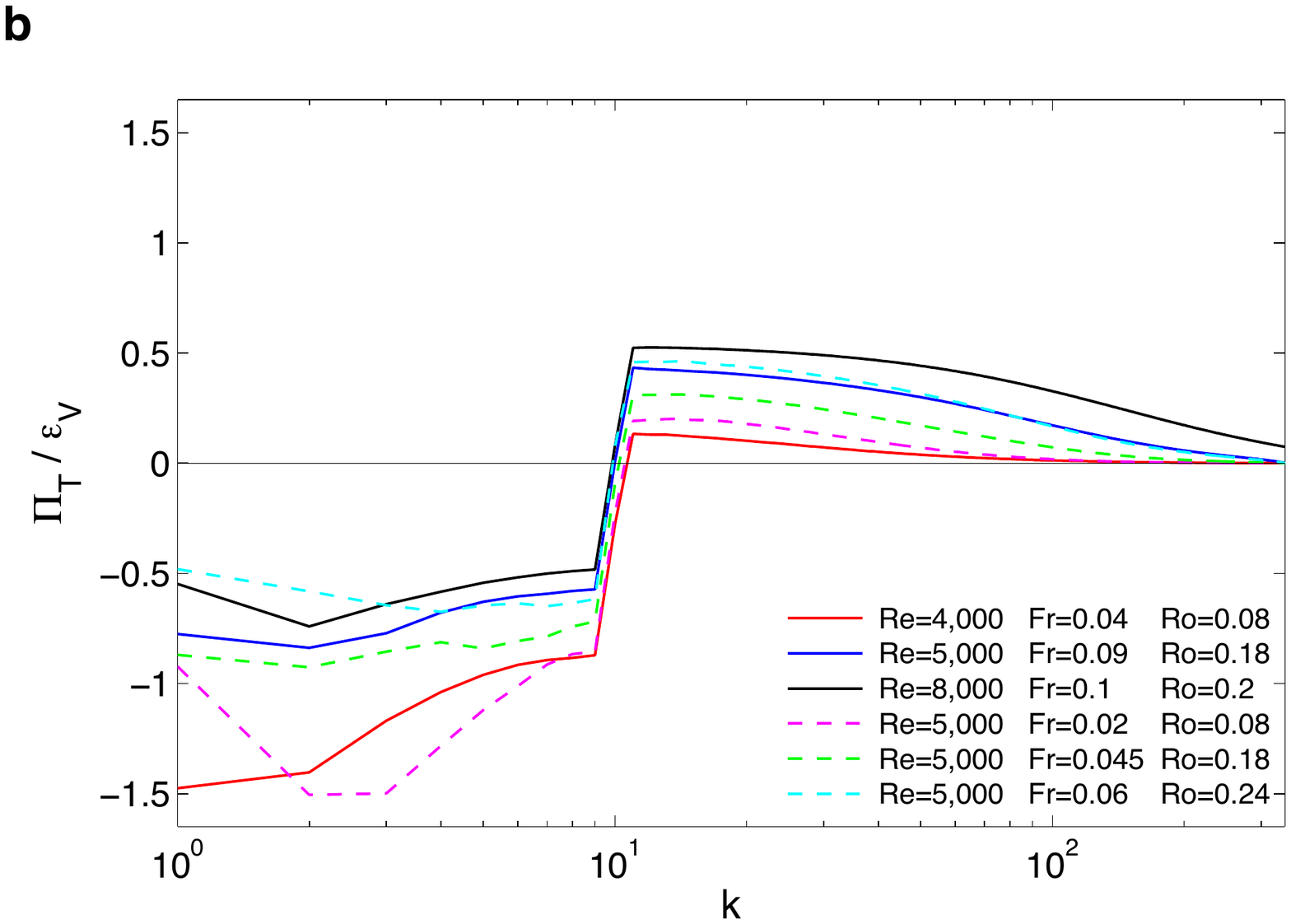}
\vskip-0.97truein
\caption{\label{2runs} 
{(Color Online.)}
{\bf (a):}
Kinetic energy spectra for run 10d (red), 10e (blue), 15a (black), 
 all with $N/f=2$ and increasing ${\cal R}_B$. The straight lines with 
different power laws are given as indications. In the bottom inset are shown the temporal evolution of the kinetic energy for the same runs 
(solid lines), together with their (scaled) dissipation (dashed lines) $5\times 2\nu \left<|\omega|^2 \right>$, with $\mathbf \omega = \nabla \times {\bf u}$ the 
vorticity. The spectra, not averaged in time, are shown at $t/{{\tau}_{NL}} \sim 22$,
whereas the peak of dissipation occurs for all the runs around 
$t/{{\tau}_{NL}} \sim 1.3$, time after which the energy starts to grow.
{\bf (b):}
Total (kinetic plus potential) energy fluxes normalized by energy input $\epsilon_V = \left< {\bf u} \cdot {\bf F} \right>$ for the same runs, as well as 
run 10a (magenta-dash), 10b (green-dash) and 10c (cyan-dash) for which  $N/f=4$.
} \end{figure*}

Results concerning scale-to-scale distribution in Fourier space are displayed in Fig.\ref{2runs} for runs with $N/f=$ 2 {and 4}, with the fluxes {$\Pi_T(k)$} (right) being averaged for 10 turn-over times after the peak of dissipation $t_p\approx 1.3$, which also marks the onset of the
inverse cascade.  All runs listed in Table \ref{tab} display a clear inverse energy cascade ($k<k_F$), with a negative flux, and with an approximate  $k^{-5/3}$ 
scaling {\cite{EPL}}, as expected from classical theory of two-dimensional (2D) turbulence \cite{rhk_montgo, ecke}. 
This inverse cascade to large scales in 2D was demonstrated using e.g. two-point closures of turbulence \cite{pouquet_75}, or more recently high-resolution numerical simulations \cite{boffetta_07}.

These runs also have a clear direct  energy cascade  % to small scales 
($k>k_F$), with a constant  positive flux. Spectral indices $\alpha$ are defined through 
$E_V(k)\sim k^{\alpha}$ where the fit is performed in the  inertial range of wavenumber, $k_F< k < k_{diss}$ with $k_{diss}\approx k_{max}$ marking the onset of
the dissipation range. These  exponents (see Table \ref{tab}), vary between {$\approx 3.99$ and $\approx 1.87$}, 
the steeper the lower $Re$ and ${\cal R}_B$. %the Reynolds and buoyancy Reynolds number. 
The shallower spectrum is  close to a Kolmogorov solution $\alpha_{Kol}=5/3$,  expected (with small intermittency corrections) once the small scales recover isotropy for high enough ${\cal R}_B$ {(see \cite{3072} for the rotating case).}

The inset in Fig. \ref{2runs} gives the temporal variation of %kinetic energy 
$E_V$ {(solid lines)} and (scaled) dissipation 
${\cal D}_V=2\nu \left< |\omega|^2 \right>$ {(dashed lines)}. 
The steady energy increase, after an initial transient, is typical of inverse cascades; 
The variation of the ratio of {inverse to direct} flux with the buoyancy Reynolds number is indicative of the increased 
effectiveness of turbulence as ${\cal R}_B$ grows. 
{One can expect this ratio to decrease as $N/f$ increases since no inverse cascade occurs in the purely stratified case \cite{EPL}.} 

Such direct cascades of energy in rotating stratified turbulence have been analyzed using theoretical closure models of turbulence \cite{galperin_10}.
Dual cascades were also found when examining AVISO altimeter data for the Kuroshio current \cite{arbic_13}, with values of $R_\Pi$ approaching those of oceanic 
data for the largest imposed turbulent (horizontal) viscosity.
Whereas these authors conclude to some ambiguity in the interpretation of their results due to the necessary filtering of the data, our DNS of the Boussinesq 
equations unambiguously show  that dual energy cascades are realistic outcomes in a geophysical setting. The higher values of $R_\Pi$ found in our runs likely reflect the fact that buoyancy is not dominant in our DNS, with $N/f \le 4$. 
{However, we note that the abyssal southern ocean at mid latitudes has $N/f$ as low as 4 or 5 and shows considerable mixing \cite{ledwell_00, heywood_02}.}

{\underline{Conclusion and discussion:}}
We have shown in this paper that a dual (direct {\sl and} inverse) constant flux energy cascade is present in rotating stratified turbulence, thereby  resolving the paradox noted by some authors (see, eg., \cite{mcwilliams_10, arbic_13}) and thus adding credence to having % displaying % justifying 
% both a large-scale geostrophic balance together with small-scale mixing in geophysical turbulence.
both geostrophic balance and anomalous transport in geophysical turbulence.
The computations clearly point out the possibility of the co-existence in the ocean and the atmosphere of  idealized large-scale dynamics dominated by quasi-geostrophic motions, together with the production of  small scales,  essential to mixing \cite{heywood_02}.

More computations and data analysis are required to categorize in a quantitative way the mixing efficiency one can expect in such flows. For 
example, the  variation of $R_\Pi$ with the relevant dimensionless parameters, is an open problem which requires huge numeral as well as observational resources. Sub-grid scale modeling of small-scale dynamics may be introduced to study this phenomenon in a parametric fashion (see e.g. \cite{pouquet_11} for  rotating flows).
However, there are some indications of a dual flux, using quasi-geostrophy \cite{arbic_13}, or in  more complex settings using a  numerical oceanic model 
applied to the California coastal current \cite{capet2008c}.
This somewhat paradoxical behavior of the energy directivity can be understood if one recalls that triadic energetic exchanges can be either  positive or 
negative, and it is a delicate balance between the two that determines the overall sign of the flux, as also found for helical flows \cite{biferale_13a}. 

Physical descriptions beyond the Boussinesq equations can be used in modeling geophysical turbulence. For example, one can consider the evaporatively-driven (as opposed to radiatively driven) configurations of stratocumulus clouds, in which case the buoyancy term is altered by the existence of a threshold (in saturation mixture fraction), leading to a nonlinear equation of state. Similar phenomena may occur in the oceans, for which there is a complex set of state relations between temperature, density and salinity which may lead to distorted isopycnal surfaces. However, using  the Boussinesq framework, it is clear that, beyond the energy cascades with  small-scale or (exclusive) large-scale constant fluxes, other -- mixed-- solutions are found that  explain how the oceanic and atmospheric systems are in quasi-geostrophic balance at large scale and {yet} have a sufficient production of small scales  leading to enhanced mixing. 

{\it We are thankful to 
B. Galperin and C. Herbert for fruitful discussions.
This work is  supported by NSF through grant CMG/1025183, and NCAR. Computations were performed at  NCAR (ASD), NSF/XSEDE
% ASC090050 \& 
TGPHY-100029 \& 110044, and  INCITE/
DOE DE-AC05-00OR22725.}

 \bibliography{ap_sept13.bib}

\begin{thebibliography}{40}%
\makeatletter
\providecommand \@ifxundefined [1]{%
 \@ifx{#1\undefined}
}%
\providecommand \@ifnum [1]{%
 \ifnum #1\expandafter \@firstoftwo
 \else \expandafter \@secondoftwo
 \fi
}%
\providecommand \@ifx [1]{%
 \ifx #1\expandafter \@firstoftwo
 \else \expandafter \@secondoftwo
 \fi
}%
\providecommand \natexlab [1]{#1}%
\providecommand \enquote  [1]{``#1''}%
\providecommand \bibnamefont  [1]{#1}%
\providecommand \bibfnamefont [1]{#1}%
\providecommand \citenamefont [1]{#1}%
\providecommand \href@noop [0]{\@secondoftwo}%
\providecommand \href [0]{\begingroup \@sanitize@url \@href}%
\providecommand \@href[1]{\@@startlink{#1}\@@href}%
\providecommand \@@href[1]{\endgroup#1\@@endlink}%
\providecommand \@sanitize@url [0]{\catcode `\\12\catcode `\$12\catcode
  `\&12\catcode `\#12\catcode `\^12\catcode `\_12\catcode `\%12\relax}%
\providecommand \@@startlink[1]{}%
\providecommand \@@endlink[0]{}%
\providecommand \url  [0]{\begingroup\@sanitize@url \@url }%
\providecommand \@url [1]{\endgroup\@href {#1}{\urlprefix }}%
\providecommand \urlprefix  [0]{URL }%
\providecommand \Eprint [0]{\href }%
\providecommand \doibase [0]{http://dx.doi.org/}%
\providecommand \selectlanguage [0]{\@gobble}%
\providecommand \bibinfo  [0]{\@secondoftwo}%
\providecommand \bibfield  [0]{\@secondoftwo}%
\providecommand \translation [1]{[#1]}%
\providecommand \BibitemOpen [0]{}%
\providecommand \bibitemStop [0]{}%
\providecommand \bibitemNoStop [0]{.\EOS\space}%
\providecommand \EOS [0]{\spacefactor3000\relax}%
\providecommand \BibitemShut  [1]{\csname bibitem#1\endcsname}%
\let\auto@bib@innerbib\@empty
%</preamble>
\bibitem [{\citenamefont {Ledwell}\ \emph {et~al.}(2000)\citenamefont
  {Ledwell}, \citenamefont {Montgomery}, \citenamefont {Polzin}, \citenamefont
  {St-Laurent}, \citenamefont {Schmitt},\ and\ \citenamefont
  {Toole}}]{ledwell_00}%
  \BibitemOpen
  \bibfield  {author} {\bibinfo {author} {\bibfnamefont {J.~R.}\ \bibnamefont
  {Ledwell}}, \bibinfo {author} {\bibfnamefont {E.~T.}\ \bibnamefont
  {Montgomery}}, \bibinfo {author} {\bibfnamefont {K.~L.}\ \bibnamefont
  {Polzin}}, \bibinfo {author} {\bibfnamefont {L.}~\bibnamefont {St-Laurent}},
  \bibinfo {author} {\bibfnamefont {R.~W.}\ \bibnamefont {Schmitt}}, \ and\
  \bibinfo {author} {\bibfnamefont {J.~M.}\ \bibnamefont {Toole}},\ }\href@noop
  {} {\bibfield  {journal} {\bibinfo  {journal} {Nature}\ }\textbf {\bibinfo
  {volume} {403}},\ \bibinfo {pages} {179} (\bibinfo {year}
  {2000})}\BibitemShut {NoStop}%
\bibitem [{\citenamefont {Vanneste}(2013)}]{vanneste_13}%
  \BibitemOpen
  \bibfield  {author} {\bibinfo {author} {\bibfnamefont {J.}~\bibnamefont
  {Vanneste}},\ }\href@noop {} {\bibfield  {journal} {\bibinfo  {journal} {Ann.
  Rev. Fluid Mech.}\ }\textbf {\bibinfo {volume} {45}},\ \bibinfo {pages} {147}
  (\bibinfo {year} {2013})}\BibitemShut {NoStop}%
\bibitem [{\citenamefont {Hoskins}\ and\ \citenamefont
  {Bretherton}(1972)}]{hoskins_72}%
  \BibitemOpen
  \bibfield  {author} {\bibinfo {author} {\bibfnamefont {B.}~\bibnamefont
  {Hoskins}}\ and\ \bibinfo {author} {\bibfnamefont {F.}~\bibnamefont
  {Bretherton}},\ }\href@noop {} {\bibfield  {journal} {\bibinfo  {journal} {J.
  Atmos. Sci.}\ }\textbf {\bibinfo {volume} {29}},\ \bibinfo {pages} {11}
  (\bibinfo {year} {1972})}\BibitemShut {NoStop}%
\bibitem [{\citenamefont {Molemaker}, \citenamefont {McWilliams},\ and\
  \citenamefont {Capet}(2010)}]{mcwilliams_10}%
  \BibitemOpen
  \bibfield  {author} {\bibinfo {author} {\bibfnamefont {M.}~\bibnamefont
  {Molemaker}}, \bibinfo {author} {\bibfnamefont {J.}~\bibnamefont
  {McWilliams}}, \ and\ \bibinfo {author} {\bibfnamefont {X.}~\bibnamefont
  {Capet}},\ }\href@noop {} {\bibfield  {journal} {\bibinfo  {journal} {J.
  Fluid Mech.}\ }\textbf {\bibinfo {volume} {{\bf 654}}},\ \bibinfo {pages}
  {35} (\bibinfo {year} {2010})}\BibitemShut {NoStop}%
\bibitem [{\citenamefont {Ivey}, \citenamefont {Winters},\ and\ \citenamefont
  {Koseff}(2008)}]{ivey_08}%
  \BibitemOpen
  \bibfield  {author} {\bibinfo {author} {\bibfnamefont {G.}~\bibnamefont
  {Ivey}}, \bibinfo {author} {\bibfnamefont {K.}~\bibnamefont {Winters}}, \
  and\ \bibinfo {author} {\bibfnamefont {J.}~\bibnamefont {Koseff}},\
  }\href@noop {} {\bibfield  {journal} {\bibinfo  {journal} {Ann. Rev. Fluid
  Mech.}\ }\textbf {\bibinfo {volume} {{\bf 40}}},\ \bibinfo {pages} {169}
  (\bibinfo {year} {2008})}\BibitemShut {NoStop}%
\bibitem [{\citenamefont {Nikurashin}, \citenamefont {Vallis},\ and\
  \citenamefont {Adcroft}(2013)}]{nikurashin_13}%
  \BibitemOpen
  \bibfield  {author} {\bibinfo {author} {\bibfnamefont {M.}~\bibnamefont
  {Nikurashin}}, \bibinfo {author} {\bibfnamefont {G.~K.}\ \bibnamefont
  {Vallis}}, \ and\ \bibinfo {author} {\bibfnamefont {A.}~\bibnamefont
  {Adcroft}},\ }\href@noop {} {\bibfield  {journal} {\bibinfo  {journal}
  {Nature Geosci.}\ }\textbf {\bibinfo {volume} {6}},\ \bibinfo {pages} {48}
  (\bibinfo {year} {2013})}\BibitemShut {NoStop}%
\bibitem [{\citenamefont {Kraichnan}\ and\ \citenamefont
  {Montgomery}(1980)}]{rhk_montgo}%
  \BibitemOpen
  \bibfield  {author} {\bibinfo {author} {\bibfnamefont {R.}~\bibnamefont
  {Kraichnan}}\ and\ \bibinfo {author} {\bibfnamefont {D.}~\bibnamefont
  {Montgomery}},\ }\href@noop {} {\bibfield  {journal} {\bibinfo  {journal}
  {Rep. Prog. Phys.}\ }\textbf {\bibinfo {volume} {43}},\ \bibinfo {pages}
  {547} (\bibinfo {year} {1980})}\BibitemShut {NoStop}%
\bibitem [{\citenamefont {Galperin}\ \emph {et~al.}(2004)\citenamefont
  {Galperin}, \citenamefont {Nakano}, \citenamefont {Huang},\ and\
  \citenamefont {Sukoriansky}}]{galperin_04}%
  \BibitemOpen
  \bibfield  {author} {\bibinfo {author} {\bibfnamefont {B.}~\bibnamefont
  {Galperin}}, \bibinfo {author} {\bibfnamefont {H.}~\bibnamefont {Nakano}},
  \bibinfo {author} {\bibfnamefont {H.-P.}\ \bibnamefont {Huang}}, \ and\
  \bibinfo {author} {\bibfnamefont {S.}~\bibnamefont {Sukoriansky}},\
  }\href@noop {} {\bibfield  {journal} {\bibinfo  {journal} {Geophys. Res.
  Lett.}\ }\textbf {\bibinfo {volume} {31}},\ \bibinfo {pages} {L13303}
  (\bibinfo {year} {2004})}\BibitemShut {NoStop}%
\bibitem [{\citenamefont {Sagaut}\ and\ \citenamefont
  {Cambon}(2008)}]{sagaut_cambon_08}%
  \BibitemOpen
  \bibfield  {author} {\bibinfo {author} {\bibfnamefont {P.}~\bibnamefont
  {Sagaut}}\ and\ \bibinfo {author} {\bibfnamefont {C.}~\bibnamefont
  {Cambon}},\ }\href@noop {} {\emph {\bibinfo {title} {Homogeneous Turbulence
  Dynamics}}}\ (\bibinfo  {publisher} {Cambridge University Press, Cambridge},\
  \bibinfo {year} {2008})\BibitemShut {NoStop}%
\bibitem [{\citenamefont {Shraiman}\ and\ \citenamefont
  {Siggia}(2000)}]{shraiman_00}%
  \BibitemOpen
  \bibfield  {author} {\bibinfo {author} {\bibfnamefont {B.~I.}\ \bibnamefont
  {Shraiman}}\ and\ \bibinfo {author} {\bibfnamefont {E.}~\bibnamefont
  {Siggia}},\ }\href@noop {} {\bibfield  {journal} {\bibinfo  {journal}
  {Nature}\ }\textbf {\bibinfo {volume} {405}},\ \bibinfo {pages} {639}
  (\bibinfo {year} {2000})}\BibitemShut {NoStop}%
\bibitem [{\citenamefont {Klein}\ and\ \citenamefont
  {Lapeyre}(2009)}]{klein_09}%
  \BibitemOpen
  \bibfield  {author} {\bibinfo {author} {\bibfnamefont {P.}~\bibnamefont
  {Klein}}\ and\ \bibinfo {author} {\bibfnamefont {G.}~\bibnamefont
  {Lapeyre}},\ }\href@noop {} {\bibfield  {journal} {\bibinfo  {journal} {Ann.
  Rev. Mar. Sci.}\ }\textbf {\bibinfo {volume} {1}},\ \bibinfo {pages} {351}
  (\bibinfo {year} {2009})}\BibitemShut {NoStop}%
\bibitem [{\citenamefont {Ferrari}\ and\ \citenamefont
  {Wunsch}(2009)}]{ferrari_09}%
  \BibitemOpen
  \bibfield  {author} {\bibinfo {author} {\bibfnamefont {R.}~\bibnamefont
  {Ferrari}}\ and\ \bibinfo {author} {\bibfnamefont {C.}~\bibnamefont
  {Wunsch}},\ }\href@noop {} {\bibfield  {journal} {\bibinfo  {journal} {Ann.
  Rev. Fluid Mech.}\ }\textbf {\bibinfo {volume} {41}},\ \bibinfo {pages} {253}
  (\bibinfo {year} {2009})}\BibitemShut {NoStop}%
\bibitem [{\citenamefont {Arbic}\ \emph {et~al.}(2013)\citenamefont {Arbic},
  \citenamefont {Polzin}, \citenamefont {Scott}, \citenamefont {Richman},\ and\
  \citenamefont {Shriver}}]{arbic_13}%
  \BibitemOpen
  \bibfield  {author} {\bibinfo {author} {\bibfnamefont {B.}~\bibnamefont
  {Arbic}}, \bibinfo {author} {\bibfnamefont {K.}~\bibnamefont {Polzin}},
  \bibinfo {author} {\bibfnamefont {R.}~\bibnamefont {Scott}}, \bibinfo
  {author} {\bibfnamefont {J.}~\bibnamefont {Richman}}, \ and\ \bibinfo
  {author} {\bibfnamefont {J.}~\bibnamefont {Shriver}},\ }\href@noop {}
  {\bibfield  {journal} {\bibinfo  {journal} {J. Phys. Oceano.}\ }\textbf
  {\bibinfo {volume} {43}},\ \bibinfo {pages} {283} (\bibinfo {year}
  {2013})}\BibitemShut {NoStop}%
\bibitem [{\citenamefont {Frisch}, \citenamefont {Lesieur},\ and\ \citenamefont
  {Sulem}(1976)}]{frisch_76}%
  \BibitemOpen
  \bibfield  {author} {\bibinfo {author} {\bibfnamefont {U.}~\bibnamefont
  {Frisch}}, \bibinfo {author} {\bibfnamefont {M.}~\bibnamefont {Lesieur}}, \
  and\ \bibinfo {author} {\bibfnamefont {P.}~\bibnamefont {Sulem}},\
  }\href@noop {} {\bibfield  {journal} {\bibinfo  {journal} {Phys. Rev. Lett.}\
  }\textbf {\bibinfo {volume} {37}},\ \bibinfo {pages} {895} (\bibinfo {year}
  {1976})}\BibitemShut {NoStop}%
\bibitem [{\citenamefont {Fournier}\ and\ \citenamefont
  {Frisch}(1978)}]{fournier_78}%
  \BibitemOpen
  \bibfield  {author} {\bibinfo {author} {\bibfnamefont {J.~D.}\ \bibnamefont
  {Fournier}}\ and\ \bibinfo {author} {\bibfnamefont {U.}~\bibnamefont
  {Frisch}},\ }\href@noop {} {\bibfield  {journal} {\bibinfo  {journal} {Phys.
  Rev. A}\ }\textbf {\bibinfo {volume} {17}},\ \bibinfo {pages} {747} (\bibinfo
  {year} {1978})}\BibitemShut {NoStop}%
\bibitem [{\citenamefont {Bell}\ and\ \citenamefont {Nelkin}(1977)}]{bell_77}%
  \BibitemOpen
  \bibfield  {author} {\bibinfo {author} {\bibfnamefont {T.}~\bibnamefont
  {Bell}}\ and\ \bibinfo {author} {\bibfnamefont {M.}~\bibnamefont {Nelkin}},\
  }\href@noop {} {\bibfield  {journal} {\bibinfo  {journal} {Phys. Fluids}\
  }\textbf {\bibinfo {volume} {20}},\ \bibinfo {pages} {345} (\bibinfo {year}
  {1977})}\BibitemShut {NoStop}%
\bibitem [{\citenamefont {Lovejoy}\ and\ \citenamefont
  {Schertzer}(2012)}]{lovejoy_12a}%
  \BibitemOpen
  \bibfield  {author} {\bibinfo {author} {\bibfnamefont {S.}~\bibnamefont
  {Lovejoy}}\ and\ \bibinfo {author} {\bibfnamefont {D.}~\bibnamefont
  {Schertzer}},\ }\href@noop {} {\bibfield  {journal} {\bibinfo  {journal}
  {Multifractal Cascades and the Emergence of Atmospheric Dynamics, Cambridge
  University Press, Cambridge}\ } (\bibinfo {year} {2012})}\BibitemShut
  {NoStop}%
\bibitem [{\citenamefont {Biferale}, \citenamefont {Musacchio},\ and\
  \citenamefont {Toschi}(2012)}]{biferale_13a}%
  \BibitemOpen
  \bibfield  {author} {\bibinfo {author} {\bibfnamefont {L.}~\bibnamefont
  {Biferale}}, \bibinfo {author} {\bibfnamefont {S.}~\bibnamefont {Musacchio}},
  \ and\ \bibinfo {author} {\bibfnamefont {F.}~\bibnamefont {Toschi}},\
  }\href@noop {} {\bibfield  {journal} {\bibinfo  {journal} {Phys. Rev. Lett.}\
  }\textbf {\bibinfo {volume} {108}},\ \bibinfo {pages} {164501} (\bibinfo
  {year} {2012})}\BibitemShut {NoStop}%
\bibitem [{\citenamefont {Celani}, \citenamefont {Musacchio},\ and\
  \citenamefont {Vincenzi}(2010)}]{celani}%
  \BibitemOpen
  \bibfield  {author} {\bibinfo {author} {\bibfnamefont {A.}~\bibnamefont
  {Celani}}, \bibinfo {author} {\bibfnamefont {S.}~\bibnamefont {Musacchio}}, \
  and\ \bibinfo {author} {\bibfnamefont {D.}~\bibnamefont {Vincenzi}},\
  }\href@noop {} {\bibfield  {journal} {\bibinfo  {journal} {Phys. Rev. Lett.}\
  }\textbf {\bibinfo {volume} {104}},\ \bibinfo {pages} {184506} (\bibinfo
  {year} {2010})}\BibitemShut {NoStop}%
\bibitem [{\citenamefont {Xia}\ \emph {et~al.}(2011)\citenamefont {Xia},
  \citenamefont {Byrne}, \citenamefont {Falkovich},\ and\ \citenamefont
  {Shats}}]{xia_11}%
  \BibitemOpen
  \bibfield  {author} {\bibinfo {author} {\bibfnamefont {H.}~\bibnamefont
  {Xia}}, \bibinfo {author} {\bibfnamefont {D.}~\bibnamefont {Byrne}}, \bibinfo
  {author} {\bibfnamefont {G.}~\bibnamefont {Falkovich}}, \ and\ \bibinfo
  {author} {\bibfnamefont {M.}~\bibnamefont {Shats}},\ }\href@noop {}
  {\bibfield  {journal} {\bibinfo  {journal} {Nature Phys.}\ }\textbf {\bibinfo
  {volume} {7}},\ \bibinfo {pages} {321} (\bibinfo {year} {2011})}\BibitemShut
  {NoStop}%
\bibitem [{\citenamefont {Davis}\ and\ \citenamefont {Yan}(2004)}]{davis_04}%
  \BibitemOpen
  \bibfield  {author} {\bibinfo {author} {\bibfnamefont {A.}~\bibnamefont
  {Davis}}\ and\ \bibinfo {author} {\bibfnamefont {X.-H.}\ \bibnamefont
  {Yan}},\ }\href@noop {} {\bibfield  {journal} {\bibinfo  {journal} {Geophys.
  Res. Lett.}\ }\textbf {\bibinfo {volume} {31}},\ \bibinfo {pages} {L17304}
  (\bibinfo {year} {2004})}\BibitemShut {NoStop}%
\bibitem [{\citenamefont {Sundkvist}\ \emph {et~al.}(2005)\citenamefont
  {Sundkvist}, \citenamefont {Krasnoselskikh}, \citenamefont {Shukla},
  \citenamefont {Vaivads}, \citenamefont {Andr\'e}, \citenamefont {Buchert},\
  and\ \citenamefont {R\`eme}}]{sundkvist_05}%
  \BibitemOpen
  \bibfield  {author} {\bibinfo {author} {\bibfnamefont {D.}~\bibnamefont
  {Sundkvist}}, \bibinfo {author} {\bibfnamefont {V.}~\bibnamefont
  {Krasnoselskikh}}, \bibinfo {author} {\bibfnamefont {P.}~\bibnamefont
  {Shukla}}, \bibinfo {author} {\bibfnamefont {A.}~\bibnamefont {Vaivads}},
  \bibinfo {author} {\bibfnamefont {M.}~\bibnamefont {Andr\'e}}, \bibinfo
  {author} {\bibfnamefont {S.}~\bibnamefont {Buchert}}, \ and\ \bibinfo
  {author} {\bibfnamefont {H.}~\bibnamefont {R\`eme}},\ }\href@noop {}
  {\bibfield  {journal} {\bibinfo  {journal} {Nature}\ }\textbf {\bibinfo
  {volume} {436}},\ \bibinfo {pages} {825} (\bibinfo {year}
  {2005})}\BibitemShut {NoStop}%
\bibitem [{\citenamefont {Pumir}\ and\ \citenamefont
  {Shraiman}(1995)}]{pumir_95}%
  \BibitemOpen
  \bibfield  {author} {\bibinfo {author} {\bibfnamefont {A.}~\bibnamefont
  {Pumir}}\ and\ \bibinfo {author} {\bibfnamefont {B.~I.}\ \bibnamefont
  {Shraiman}},\ }\href@noop {} {\bibfield  {journal} {\bibinfo  {journal}
  {Phys. Rev. Lett.}\ }\textbf {\bibinfo {volume} {75}},\ \bibinfo {pages}
  {3114} (\bibinfo {year} {1995})}\BibitemShut {NoStop}%
\bibitem [{\citenamefont {Fritts}, \citenamefont {Wang},\ and\ \citenamefont
  {Werne}(2009)}]{fritts_09a}%
  \BibitemOpen
  \bibfield  {author} {\bibinfo {author} {\bibfnamefont {D.~C.}\ \bibnamefont
  {Fritts}}, \bibinfo {author} {\bibfnamefont {L.}~\bibnamefont {Wang}}, \ and\
  \bibinfo {author} {\bibfnamefont {J.}~\bibnamefont {Werne}},\ }\href@noop {}
  {\bibfield  {journal} {\bibinfo  {journal} {Geophys. Res. Lett.}\ }\textbf
  {\bibinfo {volume} {36}},\ \bibinfo {pages} {396} (\bibinfo {year}
  {2009})}\BibitemShut {NoStop}%
\bibitem [{\citenamefont {Billant}\ and\ \citenamefont
  {Chomaz}(2001)}]{chomaz}%
  \BibitemOpen
  \bibfield  {author} {\bibinfo {author} {\bibfnamefont {P.}~\bibnamefont
  {Billant}}\ and\ \bibinfo {author} {\bibfnamefont {J.-M.}\ \bibnamefont
  {Chomaz}},\ }\href@noop {} {\bibfield  {journal} {\bibinfo  {journal} {Phys.
  Fluids}\ }\textbf {\bibinfo {volume} {13}},\ \bibinfo {pages} {1645}
  (\bibinfo {year} {2001})}\BibitemShut {NoStop}%
\bibitem [{\citenamefont {Lindborg}(2006)}]{lindborg2006}%
  \BibitemOpen
  \bibfield  {author} {\bibinfo {author} {\bibfnamefont {E.}~\bibnamefont
  {Lindborg}},\ }\href@noop {} {\bibfield  {journal} {\bibinfo  {journal} {J.
  Fluid Mech.}\ }\textbf {\bibinfo {volume} {550}},\ \bibinfo {pages} {207}
  (\bibinfo {year} {2006})}\BibitemShut {NoStop}%
\bibitem [{\citenamefont {Brethouwer}\ \emph {et~al.}(2007)\citenamefont
  {Brethouwer}, \citenamefont {Billant}, \citenamefont {Lindborg},\ and\
  \citenamefont {Chomaz}}]{brethouwer_07}%
  \BibitemOpen
  \bibfield  {author} {\bibinfo {author} {\bibfnamefont {G.}~\bibnamefont
  {Brethouwer}}, \bibinfo {author} {\bibfnamefont {P.}~\bibnamefont {Billant}},
  \bibinfo {author} {\bibfnamefont {E.}~\bibnamefont {Lindborg}}, \ and\
  \bibinfo {author} {\bibfnamefont {J.-M.}\ \bibnamefont {Chomaz}},\
  }\href@noop {} {\bibfield  {journal} {\bibinfo  {journal} {J. Fluid Mech.}\
  }\textbf {\bibinfo {volume} {585}},\ \bibinfo {pages} {343} (\bibinfo {year}
  {2007})}\BibitemShut {NoStop}%
\bibitem [{\citenamefont {Waite}\ and\ \citenamefont
  {Smolarkiewicz}(2008)}]{waite_08}%
  \BibitemOpen
  \bibfield  {author} {\bibinfo {author} {\bibfnamefont {M.}~\bibnamefont
  {Waite}}\ and\ \bibinfo {author} {\bibfnamefont {P.}~\bibnamefont
  {Smolarkiewicz}},\ }\href@noop {} {\bibfield  {journal} {\bibinfo  {journal}
  {J. Fluid Mech.}\ }\textbf {\bibinfo {volume} {606}},\ \bibinfo {pages} {239}
  (\bibinfo {year} {2008})}\BibitemShut {NoStop}%
\bibitem [{\citenamefont {Billant}\ \emph {et~al.}(2010)\citenamefont
  {Billant}, \citenamefont {Deloncle}, \citenamefont {Chomaz},\ and\
  \citenamefont {Otheguy}}]{billant_10}%
  \BibitemOpen
  \bibfield  {author} {\bibinfo {author} {\bibfnamefont {P.}~\bibnamefont
  {Billant}}, \bibinfo {author} {\bibfnamefont {A.}~\bibnamefont {Deloncle}},
  \bibinfo {author} {\bibfnamefont {J.-M.}\ \bibnamefont {Chomaz}}, \ and\
  \bibinfo {author} {\bibfnamefont {P.}~\bibnamefont {Otheguy}},\ }\href@noop
  {} {\bibfield  {journal} {\bibinfo  {journal} {J. Fluid Mech.}\ }\textbf
  {\bibinfo {volume} {660}},\ \bibinfo {pages} {396} (\bibinfo {year}
  {2010})}\BibitemShut {NoStop}%
\bibitem [{\citenamefont {Marino}\ \emph {et~al.}(2013)\citenamefont {Marino},
  \citenamefont {Mininni}, \citenamefont {Rosenberg},\ and\ \citenamefont
  {Pouquet}}]{EPL}%
  \BibitemOpen
  \bibfield  {author} {\bibinfo {author} {\bibfnamefont {R.}~\bibnamefont
  {Marino}}, \bibinfo {author} {\bibfnamefont {P.}~\bibnamefont {Mininni}},
  \bibinfo {author} {\bibfnamefont {D.}~\bibnamefont {Rosenberg}}, \ and\
  \bibinfo {author} {\bibfnamefont {A.}~\bibnamefont {Pouquet}},\ }\href@noop
  {} {\bibfield  {journal} {\bibinfo  {journal} {EuroPhys. Lett.}\ }\textbf
  {\bibinfo {volume} {102}},\ \bibinfo {pages} {44006} (\bibinfo {year}
  {2013})}\BibitemShut {NoStop}%
\bibitem [{\citenamefont {Mininni}, \citenamefont {Rosenberg},\ and\
  \citenamefont {Pouquet}(2012)}]{3072}%
  \BibitemOpen
  \bibfield  {author} {\bibinfo {author} {\bibfnamefont {P.}~\bibnamefont
  {Mininni}}, \bibinfo {author} {\bibfnamefont {D.}~\bibnamefont {Rosenberg}},
  \ and\ \bibinfo {author} {\bibfnamefont {A.}~\bibnamefont {Pouquet}},\
  }\href@noop {} {\bibfield  {journal} {\bibinfo  {journal} {J. Fluid Mech.}\
  }\textbf {\bibinfo {volume} {{\bf 699}}},\ \bibinfo {pages} {263 } (\bibinfo
  {year} {2012})}\BibitemShut {NoStop}%
\bibitem [{\citenamefont {Aluie}\ and\ \citenamefont
  {Kurien}(2011)}]{aluie_11}%
  \BibitemOpen
  \bibfield  {author} {\bibinfo {author} {\bibfnamefont {H.}~\bibnamefont
  {Aluie}}\ and\ \bibinfo {author} {\bibfnamefont {S.}~\bibnamefont {Kurien}},\
  }\href@noop {} {\bibfield  {journal} {\bibinfo  {journal} {Eur. Phys. Lett.}\
  }\textbf {\bibinfo {volume} {96}},\ \bibinfo {pages} {44006} (\bibinfo {year}
  {2011})}\BibitemShut {NoStop}%
\bibitem [{\citenamefont {Mininni}\ \emph {et~al.}(2011)\citenamefont
  {Mininni}, \citenamefont {Rosenberg}, \citenamefont {Reddy},\ and\
  \citenamefont {Pouquet}}]{hybrid2011}%
  \BibitemOpen
  \bibfield  {author} {\bibinfo {author} {\bibfnamefont {P.}~\bibnamefont
  {Mininni}}, \bibinfo {author} {\bibfnamefont {D.}~\bibnamefont {Rosenberg}},
  \bibinfo {author} {\bibfnamefont {R.}~\bibnamefont {Reddy}}, \ and\ \bibinfo
  {author} {\bibfnamefont {A.}~\bibnamefont {Pouquet}},\ }\href@noop {}
  {\bibfield  {journal} {\bibinfo  {journal} {Parallel Computing}\ }\textbf
  {\bibinfo {volume} {37}},\ \bibinfo {pages} {316} (\bibinfo {year}
  {2011})}\BibitemShut {NoStop}%
\bibitem [{\citenamefont {Boffetta}\ and\ \citenamefont {Ecke}(2012)}]{ecke}%
  \BibitemOpen
  \bibfield  {author} {\bibinfo {author} {\bibfnamefont {G.}~\bibnamefont
  {Boffetta}}\ and\ \bibinfo {author} {\bibfnamefont {R.}~\bibnamefont
  {Ecke}},\ }\href@noop {} {\bibfield  {journal} {\bibinfo  {journal} {Ann.
  Rev. Fluid Mech.}\ }\textbf {\bibinfo {volume} {44}},\ \bibinfo {pages} {427}
  (\bibinfo {year} {2012})}\BibitemShut {NoStop}%
\bibitem [{\citenamefont {Pouquet}\ \emph {et~al.}(1975)\citenamefont
  {Pouquet}, \citenamefont {Lesieur}, \citenamefont {Andr\'e},\ and\
  \citenamefont {Basdevant}}]{pouquet_75}%
  \BibitemOpen
  \bibfield  {author} {\bibinfo {author} {\bibfnamefont {A.}~\bibnamefont
  {Pouquet}}, \bibinfo {author} {\bibfnamefont {M.}~\bibnamefont {Lesieur}},
  \bibinfo {author} {\bibfnamefont {J.~C.}\ \bibnamefont {Andr\'e}}, \ and\
  \bibinfo {author} {\bibfnamefont {C.}~\bibnamefont {Basdevant}},\ }\href@noop
  {} {\bibfield  {journal} {\bibinfo  {journal} {J. Fluid Mech.}\ }\textbf
  {\bibinfo {volume} {72}},\ \bibinfo {pages} {305} (\bibinfo {year}
  {1975})}\BibitemShut {NoStop}%
\bibitem [{\citenamefont {Boffetta}(2007)}]{boffetta_07}%
  \BibitemOpen
  \bibfield  {author} {\bibinfo {author} {\bibfnamefont {G.}~\bibnamefont
  {Boffetta}},\ }\href@noop {} {\bibfield  {journal} {\bibinfo  {journal} {J.
  Fluid Mech.}\ }\textbf {\bibinfo {volume} {589}},\ \bibinfo {pages} {253}
  (\bibinfo {year} {2007})}\BibitemShut {NoStop}%
\bibitem [{\citenamefont {Galperin}\ and\ \citenamefont
  {Sukoriansky}(2010)}]{galperin_10}%
  \BibitemOpen
  \bibfield  {author} {\bibinfo {author} {\bibfnamefont {B.}~\bibnamefont
  {Galperin}}\ and\ \bibinfo {author} {\bibfnamefont {S.}~\bibnamefont
  {Sukoriansky}},\ }\href@noop {} {\bibfield  {journal} {\bibinfo  {journal}
  {Ocean Dynamics}\ }\textbf {\bibinfo {volume} {60}},\ \bibinfo {pages} {1319}
  (\bibinfo {year} {2010})}\BibitemShut {NoStop}%
\bibitem [{\citenamefont {Heywood}, \citenamefont {Garabato},\ and\
  \citenamefont {Stevens}(2002)}]{heywood_02}%
  \BibitemOpen
  \bibfield  {author} {\bibinfo {author} {\bibfnamefont {K.~J.}\ \bibnamefont
  {Heywood}}, \bibinfo {author} {\bibfnamefont {A.~N.}\ \bibnamefont
  {Garabato}}, \ and\ \bibinfo {author} {\bibfnamefont {D.}~\bibnamefont
  {Stevens}},\ }\href@noop {} {\bibfield  {journal} {\bibinfo  {journal}
  {Nature}\ }\textbf {\bibinfo {volume} {415}},\ \bibinfo {pages} {1011}
  (\bibinfo {year} {2002})}\BibitemShut {NoStop}%
\bibitem [{\citenamefont {Pouquet}\ \emph {et~al.}(2011)\citenamefont
  {Pouquet}, \citenamefont {Baerenzung}, \citenamefont {Mininni}, \citenamefont
  {Rosenberg},\ and\ \citenamefont {Thalabard}}]{pouquet_11}%
  \BibitemOpen
  \bibfield  {author} {\bibinfo {author} {\bibfnamefont {A.}~\bibnamefont
  {Pouquet}}, \bibinfo {author} {\bibfnamefont {J.}~\bibnamefont {Baerenzung}},
  \bibinfo {author} {\bibfnamefont {P.}~\bibnamefont {Mininni}}, \bibinfo
  {author} {\bibfnamefont {D.}~\bibnamefont {Rosenberg}}, \ and\ \bibinfo
  {author} {\bibfnamefont {S.}~\bibnamefont {Thalabard}},\ }\href@noop {}
  {\bibfield  {journal} {\bibinfo  {journal} {J. Phys., Conf. Series, Europ.
  Turb. Conf. Proc. ETC13, K. Bajer Ed.}\ }\textbf {\bibinfo {volume} {318}},\
  \bibinfo {pages} {042015} (\bibinfo {year} {2011})}\BibitemShut {NoStop}%
\bibitem [{\citenamefont {Capet}\ \emph {et~al.}(2008)\citenamefont {Capet},
  \citenamefont {McWilliams}, \citenamefont {Molemaker},\ and\ \citenamefont
  {Shchepetkin}}]{capet2008c}%
  \BibitemOpen
  \bibfield  {author} {\bibinfo {author} {\bibfnamefont {X.}~\bibnamefont
  {Capet}}, \bibinfo {author} {\bibfnamefont {J.}~\bibnamefont {McWilliams}},
  \bibinfo {author} {\bibfnamefont {M.}~\bibnamefont {Molemaker}}, \ and\
  \bibinfo {author} {\bibfnamefont {A.}~\bibnamefont {Shchepetkin}},\
  }\href@noop {} {\bibfield  {journal} {\bibinfo  {journal} {J. \ Phys.\
  Ocean.}\ }\textbf {\bibinfo {volume} {38}},\ \bibinfo {pages} {2256}
  (\bibinfo {year} {2008})}\BibitemShut {NoStop}%
\end{thebibliography}%

\end{document}